\documentclass{optica-article}

\journal{josab} % for journals or Optica Open

\articletype{Research Article}

\usepackage{pdfpages}

\usepackage{graphicx}
\usepackage{cite} % For references like [1-3] instead of [1, 2, 3]
\usepackage{xcolor} % Colored text
\usepackage{xr} % Cross-references across different files
\usepackage{hyperref}
\hypersetup{
    colorlinks,
    linkcolor= cyan,
    citecolor= cyan,
    urlcolor= cyan,
}

\usepackage{caption}
\newcommand{\Eqref}[1]{Eq.~\eqref{#1}}
\newcommand{\SIEqref}[1]{Eq.~(\ref*{#1})} % For suppressing hyperlinks when referencing to equations in SI
\newcommand{\Secref}[1]{Section~\ref{#1}}
\newcommand{\SISecref}[1]{Section~\ref*{#1}} % For suppressing hyperlinks when referencing to sections in SI
\newcommand{\Figref}[1]{Fig.~\ref{#1}}
\newcommand{\unit}[2]{\ensuremath{#1\,\mathrm{#2}}}

\newcommand{\abs}[1]{\left\lvert#1\right\rvert}

\newcommand{\de}{\mathrm{d}}
\DeclareMathOperator{\realPart}{Re}

\begin{document}

\title{femtoPro: Real-time linear and nonlinear optics simulations}

\author{Tobias Brixner,\authormark{1,*} Stefan Mueller,\authormark{1} Andreas Müller,\authormark{2} and Sebastian von Mammen\authormark{2}}

\address{\authormark{1}Universität Würzburg, Institut für Physikalische und Theoretische Chemie, Am Hubland, 97074 Würzburg, Germany\\
\authormark{2} Universit\"at W\"urzburg, Institut f\"ur Informatik, Am Hubland, 97074 W\"urzburg, Germany\\}

\email{\authormark{*}tobias.brixner@uni-wuerzburg.de} %% email address is required; see note below about the corresponding author designation

% use {asbstract*} to suppress the copyright line. Copyright information will be added in production

\begin{abstract*} 
Real-time optics and spectroscopy simulations ideally provide results at update rates of 120~Hz or more without any noticeable delay between changing input parameters and the calculated results. Such calculations require models of sufficient speed yet adequate level of detail in the physical approximations to contain the essential features of the simulated phenomena. We discuss a representation of femtosecond laser pulses in which fast phase oscillations due to carrier frequency and due to spatial propagation are separated out and amplitude modulations due to Gaussian beam propagation are also separated and treated explicitly. We derive simplified expressions for the spatial modulations of laser beams. Further, we derive visibility and beam-overlap factors describing multi-pulse interference. We obtain simplified expressions for radius and curvature of nonlinear signal beams in the case of fundamental beams with different convergence, different beam waist, and imperfect mutual overlap. The described model is implemented in the virtual-reality laser laboratory simulation “femtoPro,” but the derived equations can be used independently for other applications.
\end{abstract*}

%%%%%%%%%%%%%%%%%%%%%%%%%%  body  %%%%%%%%%%%%%%%%%%%%%%%%%%
\section{Introduction}
Computer simulations of optical phenomena and light--matter interaction are ubiquitous across the scientific disciplines because light fields are elementary probes of nature. In particular in the natural sciences, lasers are used as tools in the realms of frequency-resolved or time-resolved spectroscopy and microscopy applications. The particular optical setups of such techniques may require complex and precise configurations. Planning experiments and analyzing experimental data is hence generally carried out with the help of appropriate computer simulation techniques. For example, optical systems are designed using geometrical ray tracing \cite{spencer_general_1962, noauthor_list_nodate}. If more precision is required, finite-difference methods are employed for numerical approximations of wave optics \cite{garcia_de_abajo_interaction_1999, garcia_de_abajo_retarded_2002-1, oskooi_meep_2010, comsol_comsol_nodate, lumerical_high-performance_nodate, kristensen_modeling_2020}. Similarly, the simulation of light--matter interactions can be performed at various levels of accuracy and computational cost, including light-induced quantum dynamics \cite{abramavicius_coherent_2009, mitric_laser-field-induced_2009, mancal_tmancal74quantarhei_2025, rose_efficient_2021, kenneweg_qdt_2024}.

While it is generally desirable to implement procedures and algorithms as efficiently as possible for a given approach, the absolute computation time for nonlinear spectroscopy simulations is typically not the main priority because it is not critical whether results are obtained after a few seconds, minutes, or hours if the desired accuracy is reached in the end. Real-time calculations are not required under such normal-use circumstances.

We have recently introduced an immersive virtual-reality (VR) simulator of an ultrafast laser laboratory (``femtoPro'') \cite{brixner_femtopro_2023, noauthor_femtopro_nodate}. It can be used to provide practical training to students or researchers and allows them to obtain practical expertise in the fundamentals of optics, basic alignment procedures, all the way up to building and using advanced time-resolved spectroscopy setups. For this purpose, real-time simulations are required of linear and nonlinear optical phenomena. In this context, ``real time'' does not signify the native femtosecond timescale of the photoinduced dynamics, but it means that the total calculation time for a complete spectroscopy experiment should be on the order of \unit{8}{ms} on a typical VR headset such that frame update rates of \unit{120}{Hz} can be reached. Such rates reduce motion blur \cite{han_assessing_2022} and motion sickness \cite{wang_effect_2023}, improving user comfort. In other words, the goal is that the user does not notice a timing difference between carrying out any linear or nonlinear optical experiment in a real laboratory (with appropriate data acquisition software) or simulating the same setup in femtoPro. While our motivation for the present work thus stems from implementing this VR simulator, the derived equations are much more general and are applicable in other optics simulation scenarios, outside of a VR context, particularly if fast computation is desired.

Thus, the purpose of the present paper is two-fold: 1) to explain the assumptions underlying the femtoPro software such that users may understand its features and limitations, and 2) to introduce and discuss physical models that can be employed for fast optics simulations on any platform and for various purposes.

The requirement for a lag-free user experience sets a top priority on the speed of the simulations. This is different from the established simulation tools and protocols listed above. Still, the accuracy needs to be sufficient to provide a realistic simulation of the spatial and temporal propagation of ultrafast laser pulses and their interaction with matter. Some requirements are therefore: We desire to take into account Gaussian beam propagation because finite-focus-size effects are relevant to simulate beam overlap alignment procedures; laser beams may be clipped if they hit the edge of an optical element or an iris; femtosecond pulses are modified by dispersion upon propagation through materials and upon resonant interactions; and second-order processes shall be included to simulate frequency conversion, all in real time as defined above.

The underlying simulation concepts are in principle known from the vast existing scientific literature, in particular from textbooks on general optics \cite{klein_optics_1986, born_principles_1999, zinth_optik._2005, novotny_principles_2012, boyd_nonlinear_2008, saleh_fundamentals_2019} and ultrafast spectroscopy \cite{mukamel_principles_1995, diels_ultrashort_2006, trebino_frequency-resolved_2002, wollenhaupt_femtosecond_2007, weiner_ultrafast_2009, cho_two-dimensional_2009, hamm_concepts_2011-1, valkunas_molecular_2013, yuen-zhou_ultrafast_2014}. However, the real-time simulation requirement necessitates a careful balance between speed and accuracy. In another manuscript, we describe the concept, implementation, and evaluation of a dynamic graph-based algorithm that is used to find a self-consistent solution of multiple propagating and interfering laser beams \cite{muller_graph-based_2025}. The dynamic graph model maps the formal network between beam segments of laser pulses propagating freely in space as ``edges,'' and optical elements (such as mirrors or lenses) that modify the laser pulse properties as ``nodes,'' and then the algorithm is set up to minimize the number of evaluation steps. In the present work, we focus on the physical model underlying the individual calculation steps of that algorithm and derive simplified mathematical expressions that facilitate real-time optics simulations.

We start by defining our representation of the electric fields of laser pulses including the spatial modulations of laser beams resulting from geometric effects of optical elements (Section~\ref{SecLaserPulses}). Then we discuss multi-beam interference (\Secref{SecNBeamInterference}), followed by spatial and spectral--temporal modulations arising from linear and nonlinear light--matter interaction (\Secref{SecModulations}) and the graphical representation of laser beams (\Secref{SecGraphicalLaserBeamRepresentation}). Exemplary simulation results are shown in \Secref{SecSimulationResults} before we conclude in \Secref{SecConclusion}.

\section{Laser pulse propagation}
\label{SecLaserPulses}
We define and derive all relevant expressions for the description of laser pulses in great detail in \SISecref{SISecElectricField} of Supplementary Information (SI) and only quote the main results here. Indeed, we highly recommend reading the entire \SISecref{SISecLaserPulsePropagationAndModificationByOpticalElements} of the SI before continuing with the main article here because all the relevant quantities and essential transformations are set up in the SI.

As a result of the derivations, we find that the electric field of a laser pulse at position $\mathbf{r}$ as a function of spatial coordinates $x$, $y$ and $z$ and of time $t$ is given by
\begin{equation}
\label{EqEPropXYZT}
  E_{\text{prop}}^+(x,y,z,t)=\sqrt{\frac{2}{\pi}}\frac{1}{w(z)}
  \exp\left[-\frac{(x-x_0)^2+(y-y_0)^2}{w^2(z)}\right]
  e^{i\mathbf{k}\cdot\mathbf{r}}
  \sqrt{\frac{S}{\delta t}}
  \tilde{E}_t(j)e^{-i\omega_0(t-T)}
\end{equation}
according to \SIEqref{SIEqEPropXYZT} in the SI, where the ``$+$'' superscript in $E_{\rm{prop}}^+$ indicates that this is a complex-valued representation resulting from the positive-frequency part of the full field only and the subscript ``prop'' indicates that the pulse has propagated through space along the $\hat{\mathbf{z}}$ direction requiring the propagation time $T$. The beam radius at each $z$ position is given by $w(z)$ [\SIEqref{SIEqBeamRadiusEvolution}] and lateral translations of the beam along $\hat{\mathbf{x}}$ and $\hat{\mathbf{y}}$ by $x_0$ and $y_0$, respectively. Further, $\mathbf{k}$ denotes the wave vector, $\omega_0$ the center frequency, $\delta t$ the sampling step size in time domain, $S$ a scaling factor proportional to pulse energy, and $\tilde{E}_t(j)$ the numerical array representation of the complex-valued temporal envelope (including amplitude and phase terms) for individual sampling points that are indexed by $j$.

Analogously, we obtain the temporally propagated spatial--spectral field from \SIEqref{SIEqEPropXYZOmega},
\begin{equation}
\label{EqEPropXYZOmega}
  E_{\textrm{prop}}^+(x,y,z,\omega)=\sqrt{\frac{2}{\pi}}\frac{1}{w(z)}
  \exp\left[-\frac{(x-x_0)^2+(y-y_0)^2}{w^2(z)}\right]
  e^{i\mathbf{k}\cdot\mathbf{r}}
  \sqrt{\frac{S}{\delta\omega}}
  \tilde{E}_\omega(j)e^{i\omega T},
\end{equation}
with analogous definitions, and where $\delta\omega$ describes the sampling step size in frequency domain.

We now discuss how to describe the propagation of laser beams in the presence of finite apertures. Apertures lead to the clipping of laser beams. Recently, analytical propagation formulas have been derived for truncated Gaussian beams \cite{worku_propagation_2019}. Here we ignore diffraction for reasons of computational speed. If we removed the Gaussian beam properties~\eqref{EqEPropXYZOmega} at the same time, that would correspond to a transition from wave optics to the limit of geometrical optics. The latter is not sufficient, however, if we want to describe interference phenomena and allow for the quantitative treatment of spatial beam overlap effects in nonlinear optical phenomena. Thus, in the present work we suggest an intermediate regime where we treat some phenomena in the limit of geometrical optics and some phenomena using wave optics.

Concerning apertures, we select a treatment by geometrical optics as discussed in detail in \SISecref{SISecFiniteAperture} of the SI. This has the advantage that we do not have to follow various diffracted beamlets that might propagate in various different directions, in particular after interaction with subsequent optical elements. Instead, we can continue to describe the transmitted beam as a single entity with one limited set of parameters. The approximation of geometrical optics entails that our model cannot be used to simulate diffraction gratings. On the other hand, this strongly reduces complexity and facilitates real-time simulations.

We seek a representation that is as faithful to reality as possible under the given model assumptions. Thus, a laser beam emerging from any open aperture should not have a cross section that extends beyond the hard limits of the aperture. Such an aperture could for example be an iris that is opened or closed by users either to clip laser beams on purpose or to act as alignment tools. The laser beam then would be centered onto at least two subsequent irises, and after the alignment procedure is completed, the irises would be opened to let the full beam pass. In that case, then, the final beam propagation would again not be limited by diffraction for the ``real'' experiment, justifying the limit of geometrical optics, while the clipping effects will be captured at least qualitatively or semi-quantitatively during the simulation of alignment procedures. As a second possibility, finite-aperture effects become relevant when a laser beam is larger than the optical element it hits, or when it hits the element at its edge. This might happen for misaligned mirrors or lenses that are hit by the laser at their edges instead of in the center. While such situations should be simulated at least qualitatively in a real-time optics education scenario, they should typically not arise in a final, correctly set up experiment. Thus, again, it seems justified to employ a simplified geometrical optics treatment.

For treating the transmission through optics, we have to consider finite incidence angles. For example, mirrors are routinely employed at $\approx 45^{\circ}$ incidence angle to deflect beams, and anyway we cannot avoid that the beam hits a general optical element (GOE) at an arbitrary angle, so the model needs to deal with all situations. Thus we may not assume the paraxial limit, and we employ vector calculus for an analysis independent of any particular chosen coordinate system.

In the case of an unclipped beam transmitted through a GOE with focal length $f$, a beam with normalized incident wave vector $\hat{\mathbf{k}}_{\text{in}}$ that hits the GOE at a displacement $d$ with respect to the GOE center will exit the element in a new direction given by the normalized outgoing wave vector $\hat{\mathbf{k}}_{\text{in}}$, where these quantities are related by \cite{gatland_thin_2002}
\begin{equation}
\label{EqDirectionChangeGOELaserCompletelyWithinAperture}
  \hat{\mathbf{k}}_{\text{out}}=\frac{\hat{\mathbf{k}}_{\text{in}}-\frac{\mathbf{d}}{f}}
  {\abs{\hat{\mathbf{k}}_{\text{in}}-\frac{\mathbf{d}}{f}}}
\end{equation}
as derived in \SIEqref{SIEqDirectionChangeGOELaserCompletelyWithinAperture}. The beam radius projected onto the plane of the GOE remains the same because there is no clipping.

In the case of partial beam clipping, the quantity $\mathbf{d}$ in \Eqref{EqDirectionChangeGOELaserCompletelyWithinAperture} needs to be replaced by an effective quantity $\mathbf{p}$ as in \SIEqref{SIEqDefinitionP} that points from the GOE center to the center of the emerging clipped beam, leading to a modified direction of the outgoing beam according to \SIEqref{SIEqDirectionChangeGOELaserPartiallyWithinAperture}.

In the case that the beam is fully clipped, the center of the transmitted beam will be at the center of the GOE even if the incident beam was laterally displaced. This is because we consider only circularly symmetric apertures such that emerging beams that are fully defined by the aperture will always be centered. In that case, the outgoing beam direction is the same as the ingoing beam direction according to \SIEqref{SIEqKOutCompleteClipping}, which can be rationalized because the direction of a ``center beam'' through a lens is unaffected by any focal length.

Finally, we need to consider the power transmission through a circular aperture. In \SISecref{SISecFiniteAperture} of the SI, we assume without loss of generality that the beam is displaced along a GOE-local $\hat{\mathbf{x}}$ coordinate with respect to the center of the partially closed GOE aperture. Then we have to solve \SIEqref{SIEqPowerTransmission},
\begin{equation}\label{EqPowerTransmission}
  P_\text{out}=I_0\underset{(x^2+y^2\leq a^2)}{\int\de x \int\de y}\,\exp\left[-2\frac{(x-d)^2}{w^2}-2\frac{y^2}{w^2}\right],
\end{equation}
where the output power $P_\text{out}$ is obtained from the on-axis incident beam intensity $I_0$, the incident beam radius $w$, the displacement $d$, and the aperture radius $a$ through integration of the spatial Gaussian beam intensity distribution over the circular opening, which is performed numerically with pre-calculated loop-up tables to speed up the computation. In the case of a straight edge instead of a circular aperture, the power transmission is obtained in closed form in \SIEqref{SIEqTransmissionStraightEdge} via the complementary error function in \SIEqref{SIEqComplementaryErrorFunction}.

\section{\textit{N}-beam interference}
\label{SecNBeamInterference}
We now analyze the interference of $N$ laser pulses, $E_{k,\text{prop}}^+(\mathbf{r},\omega)$, $k=\{1,\ldots,N\}$, each given by \Eqref{EqEPropXYZOmega}, on a plane such as a detector. For this purpose, we evaluate the spectral power $P_\omega$, as detected by a spectrometer, by integrating the absolute magnitude squared of the total field over transverse coordinates \cite{brixner_femtopro_2023},
\begin{align}
\label{}
  P_\omega(j)=&\int_{-\infty}^{\infty}\int_{-\infty}^{\infty}
  \abs{\sum_{k=1}^N E_{k,\text{prop}}^+(\mathbf{r},\omega)}^2
  \,\de x\,\de y\\
  =&\sum_{k=1}^N \frac{S_k}{\delta\omega}\abs{\tilde{E}_{k,\omega}(j)}^2
  +\sum_{k=2}^N\sum_{l=1}^{k-1}
  2\realPart\left\{\eta_{k,l}\frac{\sqrt{S_k S_l}}{\delta\omega}
  \tilde{E}_{k,\omega}(j)\tilde{E}_{l,\omega}^*(j)
  e^{i\omega(T_k-T_l)}\right\}.\label{EqSpectralInterferencePower}
\end{align}

Here we derive the factor $\eta_{k,l}$ regulating interference visibility between two fields with indices $k$ and $l$. Note that we do not want to describe the spatially resolved visibility of interference fringes, but their integrated effect when evaluating a finite detector area. Then, interference is noticeable if the spatial fringe spacing is large enough such that it does not average out when integrating over all fringes. Considering the field definition in \Eqref{EqEPropXYZOmega}, we have to evaluate
\begin{equation}
\label{}
  \begin{split}
      \eta_{k,l}=\,&\frac{2}{\pi w_k w_l}
      \int_{-\infty}^{\infty}\int_{-\infty}^{\infty}
      \exp\left[-\frac{(x-x_k)^2+(y-y_k)^2}{w_k^2}\right]
      \exp\left[-\frac{(x-x_l)^2+(y-y_l)^2}{w_l^2}\right]\\
      &\times e^{i\Delta\mathbf{k}\cdot\mathbf{r}}
      \,\de x\,\de y
  \end{split}
\end{equation}
with the wave-vector mismatch
\begin{equation}
\label{}
  \Delta\mathbf{k}=\mathbf{k}_k-\mathbf{k}_l.
\end{equation}

We proceed in the coordinate system in which the $\hat{\mathbf{z}}$ axis is parallel to the average of the incident wave vectors. As will be derived in the SI, interference is visible only for small phase mismatch, and thus the approximation is valid that the beam positions, $(x_k,y_k)$ and $(x_l,y_l)$, and radii, $w_k$ and $w_l$, need not be transformed because the average direction is almost identical to either of the incident directions. The actual incidence angle onto the detector plane is not decisive in first order because the underlying spatial interference pattern is retained, and thus integration over the tilted pattern provides a comparable total power as integration over the normal-incidence pattern.

In \SISecref{SISecOverlapOfTwoDimensionalGaussianFunctions} of the SI, we show that the product of two two-dimensional Gaussian cross-sections is another Gaussian cross-section, and we derive its resulting product width parameters $\alpha_\text{p}$ and $\beta_\text{p}$, transverse position $(x_\text{p},y_\text{p})$, and product amplitude $A_\text{p}$ (see SI for definitions). Using those results for unity incident amplitudes, we have to evaluate
\begin{equation}
\label{EqVisibilityIntegral}
  \eta_{k,l}
  =\frac{2A_{\text{p}}}{\pi w_k w_l}
  \int_{-\infty}^{\infty}\int_{-\infty}^{\infty}
  \exp\left[-\alpha_{\text{p}}(x-x_{\text{p}})^2-\beta_{\text{p}}(y-y_{\text{p}})^2\right]
  e^{i(\Delta k_x x+\Delta k_y y)}\,\de x\,\de y.
\end{equation}
This integral is solved in \SISecref{SISecInterferenceVisibilityFactor} of the SI, leading to the already reported result \cite{brixner_femtopro_2023}
\begin{equation}
\label{EqVisibilityFinalResult}
  \eta_{k,l}=\frac{2w_k w_l}{w_k^2+w_l^2}
  \exp\left[-\frac{(x_k-x_l)^2+(y_k-y_l)^2}{w_k^2+w_l^2}\right]
  e^{i(\Delta k_x x_{\text{p}}+\Delta k_y y_{\text{p}})}
  \exp\left[-\frac{\Delta k_x^2+\Delta k_y^2}
  {2\left(\frac{1}{w_k^2}+\frac{1}{w_l^2}\right)}\right].
\end{equation}

Note that interference visibility in the literature is often treated in terms of the coherence functions of the light fields. This is taken into account in \Eqref{EqSpectralInterferencePower} by the specific properties of the spectral envelope functions. We discuss a frequency-resolved version of interference in which coherence pertains for arbitrarily long differences in propagation distance (by definition of a spectral field). For the spectrally integrated result, one then obtains the usual limit of optical coherence and interference visibility that disappears when fields are separated longitudinally by a distance larger than their coherence length (or, for bandwidth-limited ultrashort pulses, by delay times larger than their pulse duration).

\section{Modulations from light--matter interaction}
\label{SecModulations}
\subsection{First-order response}
\label{SecFirstOrderNonresonantResponse}
The electric-field modifications by linear response of matter is most conveniently implemented by multiplying the incident frequency-domain field, $E_{\text{in}}(\omega)$, with a frequency-domain linear modulation function, $M(\omega)$, to get the output field,
\begin{equation}
\label{EqGeneralModulationFunction}
  E_{\text{out}}(\omega)=M(\omega)E_{\text{in}}(\omega),
\end{equation}
after transmission through the material \cite{diels_ultrashort_2006}. In the algorithmic realization, we use, equivalently, the $\tilde{E}_\omega(j)$ array.
For treating non-resonant dispersion in matter, we follow the standard convention to perform a Taylor expansion of the dispersive spectral phase $\Phi_{\text{disp}}(\omega)$ in \SISecref{SISecTaylorExpansionOfDispersivePhase} of the SI, resulting in Taylor coefficients $b_{j,\text{disp}}$ of $j$th order.

Considering the effect on $\Phi_{\text{disp}}(\omega)$ of the zeroth- and first-order Taylor coefficients explicitly and rearranging, we get
\begin{equation}
\label{}
  b_0+b_1(\omega-\omega_0)=\frac{L\omega_0(n_0-n_{\text{gr}})}{c}+\frac{L n_{\text{gr}}}{c}\omega
\end{equation}
for propagation through a medium of length $L$ with refractive index $n_0$ and group index $n_\text{gr}$, wherein the first term is a frequency-uniform phase and the second term is linear in $\omega$. Note that we treat the propagation time $T$, arising from the geometry of free-space propagation, separately according to \SIEqref{SIEqEOmegaProp} from the SI. Thus, when transmitting through a material of thickness $L$, we have to replace $T$ with
\begin{equation}
\label{EqTTotal}
  T_{\text{total}}=T+\frac{L}{c}(n_{\text{gr}}-1)
\end{equation}
and apply only the remaining terms $b_{2,\text{disp}}$ and $b_{3,\text{disp}}$ in the non-resonant part of the linear modulation function, $M_{\text{non-resonant}}(\omega)$. Note that concerning the geometry, we work in the limit of infinitesimally thin GOEs with one central ``active plane,'' so that the incident and outgoing beams have already been assumed to travel in vacuum up to and starting from that plane. Thus, we need to subtract $L/c$ from the (modified) propagation time because it has already been accounted for, leading to the term ``$-1$'' in \Eqref{EqTTotal}. In the case of reflection, group-delay dispersion or a third-order phase coefficient may be provided directly to describe, e.g., chirped mirrors.

We treat the case of first-order resonant response, on the example of a molecular sample with an electronic transition coupled to vibrational modes, in \SISecref{SISecFirstOrderResonantResponse} of the SI.

\subsection{Second-order response}
\label{SecSecondOrderResponse}
We describe the non-resonant second-order nonlinear generation of a signal field, $E_{\text{s}}(t)$, in the approximation of being proportional to the square of the sum of incident fields, $E(t)$. Considering two incident fields $E_1(t)$ and $E_2(t)$ with $E(t)=E_1(t)+E_2(t)$ and complex envelopes $\tilde{E}_1(t)$ and $\tilde{E}_2(t)$, this leads to the explicit terms
\begin{equation}
\label{EqSecondOrderTermsMultipliedOut}
\begin{split}
  E^2(t)
  &=\tilde{E}_1^2(t)e^{i(2\mathbf{k}_1\cdot\mathbf{r}-2\omega_1t)}+\text{c.c.}
  +\tilde{E}_2^2(t)e^{i(2\mathbf{k}_2\cdot\mathbf{r}-2\omega_2t)}+\text{c.c.}\\
  &+2\tilde{E}_1(t)\tilde{E}_2(t)
    e^{i[(\mathbf{k}_1+\mathbf{k}_2)\cdot\mathbf{r}-(\omega_1+\omega_2)t]}+\text{c.c.}\\
  &+2\tilde{E}_1(t)\tilde{E}_2^*(t)
    e^{i[(\mathbf{k}_1-\mathbf{k}_2)\cdot\mathbf{r}-(\omega_1-\omega_2)t]}+\text{c.c.}\\
  &+ 2\abs{\tilde{E}_1(t)}^2 + 2\abs{\tilde{E}_2(t)}^2,
\end{split}
\end{equation}
where ``c.c.'' indicates the complex conjugate of the previous term. This is the level of treatment often found in didactic textbooks on second-order response. However, we need to consider also the spatial properties of the beams if we want to simulate and understand the effect of beam alignment on nonlinear phenomena. In \SISecref{SISecDerivationSecondOrderResponse} of the SI, we derive the explicit equations in terms of the computationally stored complex envelope arrays $\tilde{E}_t(j)$ and associated parameters, we take into account the spatial beam profile, and we determine the missing proportionality factor between $E_{\text{s}}(t)$ and $E^2(t)$.

The correct description of the nonlinear signal field further requires to calculate its beam radius and the radius of curvature, depending on the incident fields. These relations are derived in \SISecref{SISecNonlinearSignalBeamCurvature} of the SI. 

When new signal fields are generated according to second-order or third-order response and no energy is absorbed in the nonlinear crystal or sample, the total energy summed up over all laser beams should be conserved. In a fully self-consistent solution of Maxwell's equations, this would arise automatically. Our simplified treatment calls for manual adjustment of fundamental pulse energies. Otherwise, it might happen that the total energy of all output beams after a nonlinear GOE exceeds the total energy of all incident beams. We arrive at the correct energy, contributed to the new beam by each fundamental, by analyzing the relevant number of photons for each microscopic signal-generation process as derived in \SISecref{SISecReductionOfFundamentalPulseEnergiesUponNonlinearSignalGeneration} of the SI. 

\section{Graphical laser beam representation}
\label{SecGraphicalLaserBeamRepresentation}
Simulating linear and nonlinear optical phenomena leads to laser beams propagating through space with parameters as determined in \Secref{SecLaserPulses} and \Secref{SecModulations}. If one only seeks numerical results, then the calculation is already complete because the obtained parameters provide a full characterization. Often it is desired, however, to represent the laser beams graphically without compromising the real-time simulation. The details of graphical visualization of laser beams will depend on the particular software implementation, which is beyond the scope of the present work. Instead, we discuss some general points that are relevant for any such three-dimensional modeling that is sufficiently fast and realistic.

In \SISecref{SISecGaussianBeamScaling} of the SI, we show that any Gaussian beam can be derived from the general shape of a ``reference beam'' by suitable scaling. Thus, it is not necessary to re-calculate, point by point, the spatial envelopes of a beam for any new set of beam parameters, but one can simply generate a Gaussian model once, and then scale it accordingly. Let us assume we have available a representation of a Gaussian ``reference beam'' that is given by its beam radius $\tilde{w}(z)$ at position $z$ along the propagation direction. From that reference, we can obtain the representation of any new Gaussian beam, $w(z)$, according to \SIEqref{SIEqMeshTransformation},
\begin{equation}
  w(z+z_0)=b^2\tilde{w}\left(\frac{z}{b}\right),
\end{equation}
by applying a scaling factor $b^2$ along the transverse coordinates ($x$ and $y$) and a scaling factor $b$ along the longitudinal propagation coordinate $z$. The result is then shifted by $-z_0$, along the positive propagation direction, to obtain the correct beam-waist position. The scaling factor
\begin{equation}
\label{}
  b=\frac{z_{\text{R}}}{\tilde{z}_{\text{R}}}
\end{equation}
is the ratio of the Rayleigh lengths of new beam and reference beam [\SIEqref{SIEqScalingFactorRayleighLengths}]. For representing finite-length beams, one can ``clip'' the resulting object appropriately at its end positions such that only the relevant section (e.g., between two mirrors) is displayed.

We now discuss how such a reference shape can be prepared efficiently. In three-dimensional computer graphics, solid objects are generally modeled as polygon meshes, i.e., a set of points (``vertices'') in three-dimensional space that are connected by straight lines (``edges'') that in turn form small planes (``faces''). The collection of vertices and edges is sought such that the resulting set of faces approximates the real surface of the given object.

In practice, striving for good performance in real-time rendering applications, one seeks to keep the number of vertices as small as possible to reduce calculation time, yet still capturing the essential shape of the underlying object as accurately as possible. In \SISecref{SISecOptimalVertexDistribution} of the SI, we find the minimum number of vertices and their positions that can be used to represent a Gaussian laser beam for a given maximum allowed error. That error is defined as the relative length difference between a curve segment following the real Gaussian shape and a straight edge, connecting the same two neighboring vertices. In particular, we found that the spatial envelope of any Gaussian beam can be visualized, at a \unit{1}{\%} error level, in three-dimensional space using a polygonal mesh with just five vertices connected by straight edges along the propagation direction to describe the hyperbolic beam evolution and $13$ vertices along the circumference to approximate the circular symmetry.

Lastly, let us discuss the perceived color of a laser beam that is relevant in realistic graphical representations. When a laser pulse is scattered off a surface, an observer can perceive the spot with a certain color and a certain luminance (``brightness'') that depends on the pulse spectrum, energy, beam radius, and surface properties. For a realistic visualization of such a cross section or of the full laser-beam mesh discussed in Sections \ref*{SISecGaussianBeamScaling} and \ref*{SISecOptimalVertexDistribution} of the SI, we assign a corresponding computer-graphics color code.

In \SISecref{SISecColorPerception} of the SI, we use a ``standard observer'' model of average human color perception as defined by the ``Commission Internationale de l'\'{E}clairage'' (CIE). We use the XYZ standard in its most recent implementation \cite{stockman_cone_2019} to get the XYZ color values for the given laser spectrum. This color code is subsequently transformed to an RGB representation that can be displayed in computer graphics. In addition, one can use the ``alpha'' channel A in the RGBA system (typically representing transparency or opacity) to define luminance, as shown in \SIEqref{SIEqLuminance}. This luminance is proportional to the pulse-energy scaling factor $S$, meaning a laser beam with higher energy will appear brighter.

Since we account for the full laser spectrum in the color calculation, this automatically leads to the effect that very short pulses (nearing a super-continuum) will appear more ``white'' compared to narrowband spectra. An adequate representation of the laser color can thus serve a useful function by providing feedback in real-time optics simulations.

\section{Simulation results}
\label{SecSimulationResults}
\begin{figure}
\centering
\includegraphics[width=.9\linewidth]{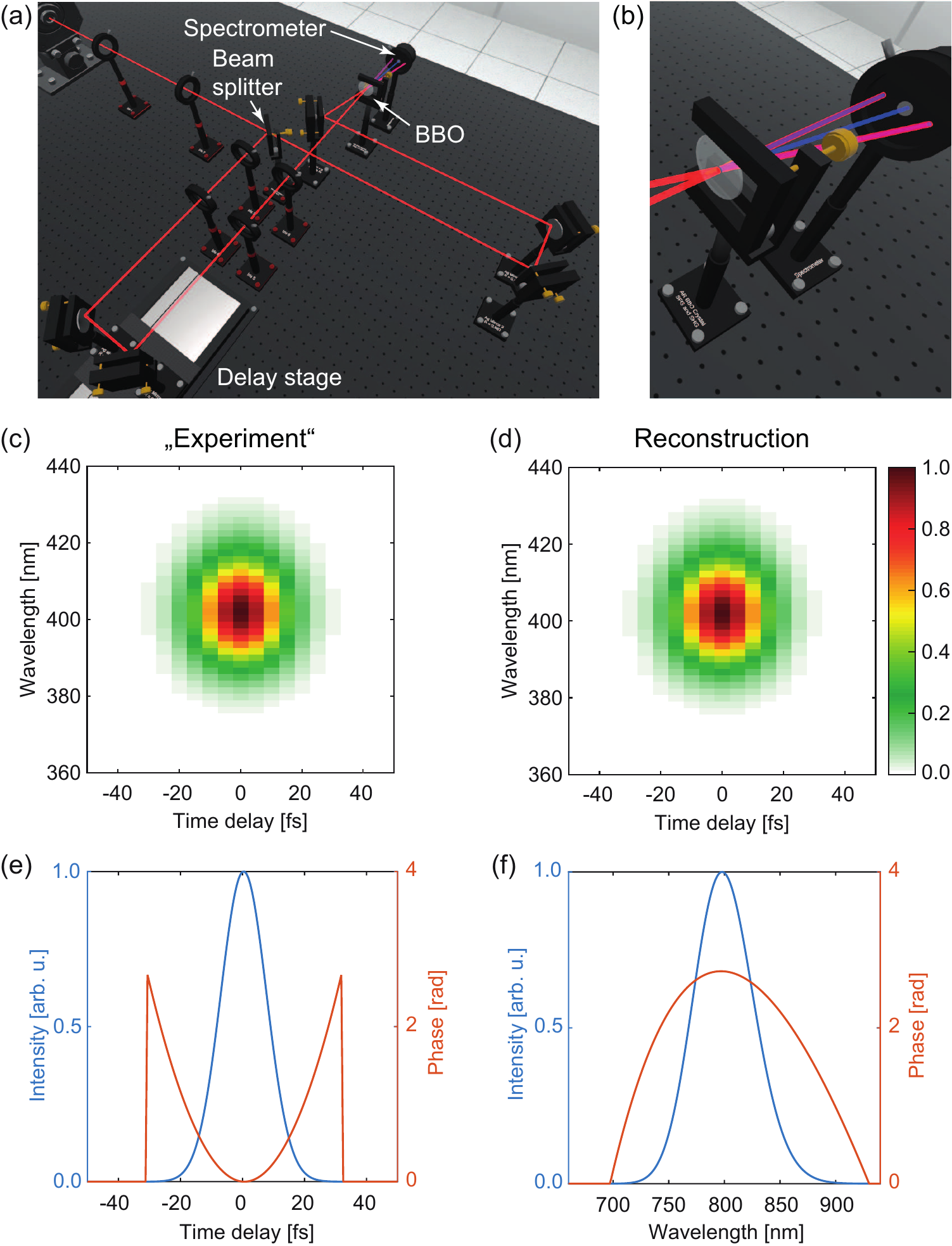}
\caption{Frequency-resolved optical gating (FROG) in VR. (a) Screenshot from femtoPro showing a Mach--Zehnder-type interferometer with second-harmonic generation (SHG) using a $\beta$-barium borate (BBO) crystal. (b) Assuming perfect phase matching, the SHG beam in the direction of the superposition of the wave vectors of the two incident beams is detected on a spectrometer. (c) ``Experimental'' FROG trace obtained in VR. (d) Reconstructed FROG trace. (e) Reconstructed temporal intensity and phase. (f) Reconstructed spectral intensity and phase.}
\label{FigSimulationResults}
\end{figure}
Having introduced our model for linear and nonlinear optics simulations, we now show exemplary simulation results. For that purpose, we consider a setup for a well-known pulse-characterization method, second-harmonic generation (SHG) frequency-resolved optical gating (FROG) \cite{delong_frequency-resolved_1994,trebino_frequency-resolved_2002}. The setup as shown in \Figref{FigSimulationResults}(a) is based on a Mach--Zehnder-type interferometer. Note that, to achieve perfect interferometer alignment, the GOEs shown in \Figref{FigSimulationResults}(a) were placed programmatically by setting numerical position values rather than by placing them ``by hand'' in VR. The laser emitter in the top-left corner of \Figref{FigSimulationResults}(a) generates Gaussian pulses with a center wavelength of 800~nm and $\tau_\text{p}=15$~fs duration [intensity full width at half maximum as in \SIEqref{SIEqNonNormalizedTemporalEnvelope}]. After passing an idealized beam splitter without dispersion, the reflected beam with wave vector $\mathbf{k}_1$ is delayed using a delay stage which can be controlled ``remotely'' through a virtual computer interface within the VR environment. The transmitted beam with wave vector $\mathbf{k}_2$ is noncollinearly overlapped with the reflected beam in a GOE that has the material parameters of a $\beta$-barium borate (BBO) crystal. Under the assumption of perfect phase matching, the second-order response of the BBO was modeled to be ``complete,'' i.e., the second harmonics of the two individual beams in $2\mathbf{k}_1$ and $2\mathbf{k}_2$ direction are generated in addition to the ``collaborative'' SHG in $\mathbf{k}_1 + \mathbf{k}_2$ direction [\Figref{FigSimulationResults}(b)]. The latter was detected by a spectrometer. In the scan procedure in VR, the delay stage was incremented from $-100$~fs to $+100$~fs in steps of $0.2$~fs. Rather than directly calculating the resulting FROG trace from an idealized ``FROG signal equation,'' the data was acquired analogously to reality, i.e., after moving the delay stage to a new step position, the SHG spectrum resulting from the interference of $\mathbf{k}_1$ and $\mathbf{k}_2$ in the BBO was calculated according to \Secref{SecSecondOrderResponse} before moving to the next step position. 

The entire scan took \unit{45}{s}, and the consistently high and stable frame rate allowed real-time monitoring of changes in the fringe pattern detected by the spectrometer at each scanning step. The thus-obtained ``experimental'' FROG trace shown in \Figref{FigSimulationResults}(c) was evaluated using a commercial FROG analysis program \cite{delong_frog_2004}, yielding the reconstruction results shown in \Figref{FigSimulationResults}(d--f).

Using a grid size of 64, the minimum FROG error was $2.5 \times 10^{-4}$. Since simulated data have been used, the reconstruction quality is excellent [compare \Figref{FigSimulationResults}(c) and \Figref{FigSimulationResults}(d)]. For a more immersive experience, artificial noise could be added to the acquired data. The slight up-chirp observed in the reconstructed electric field is a result of phase accumulation by propagation through air over the distance from the laser emitter to the BBO, leading to a reconstructed pulse duration of $17.4$~fs. Note that we use the approximation of a thin BBO, i.e., the SHG signal is calculated at the front surface of the BBO crystal such that further dispersion introduced by the finite thickness of the BBO is not taken into account.

\section{Conclusion}
\label{SecConclusion}
In the present work, we have described the development of a model for real-time simulation of linear and nonlinear optical phenomena and spectroscopy. The term ``real time'' means that we endeavor to find such a level of approximation that allows calculation times of complete experimental configurations on the order of \unit{8}{ms} on consumer-grade hardware. We chose to implement a mixture of geometrical optics and Gaussian wave optics because the former offers computational speed and simplicity, while the latter provides the correct evolution of beam radius and beam curvature during propagation. A finite beam radius is in turn important to describe effects of beam overlap in interference phenomena and nonlinear signal generation. 

We summarize in Table \ref{TableFeaturesAndLimitations} basic features and consequences of the physical model of the present work. This list is not meant to be exhaustive but rather to capture essential points for a quick overview.

\begin{table}[t]
\centering
\caption{Features and limitations of physical model.}
\label{TableFeaturesAndLimitations}
\begin{tabular}{l|l|l}
  \hline\hline
  \textbf{Property} & \textbf{Choice} & \textbf{Consequence} \\
  \hline
  Beam profile & Gaussian & Finite beam overlap included \\
  Beam quality & $M^2$ factor & Real beam divergence approximated \\
  Beam symmetry & Radially symmetric & No astigmatism\\
  Beam clipping & No diffraction & Geometrical effects included; no gratings\\
  Beam focusing & Ideal & No chromatic or spherical aberration\\
  Polarization direction & Scalar electric fields & No birefringence; no polarization\\
  Electric field & Discrete sampling & Spatial--temporal interference included\\
  First-order response & Response function & Dispersion and absorption included\\
  Second-order response & Instantaneous & Pulse-shape effects included\\
  \hline\hline
\end{tabular}
\end{table}

We provided expressions that can be used to computationally process frequency- and time-sampled electric field evolutions of femtosecond laser pulses. We removed fast phase oscillations due to the carrier frequency (from the time-domain field) and due to propagation (from the frequency-domain field) to facilitate numerical stability. In addition, we separated the pulse energy from the amplitude changes that occur due to beam-radius evolution.

We derived equations that can be used to calculate geometric beam parameters after transmission of laser beams through finite circular apertures, ignoring astigmatism and diffraction. The obtained approximations are straightforward to apply and ensure that geometric laser-beam cross sections are always contained within the GOE apertures from which they emerge. The energy throughput was determined from the numerical evaluation of a Gaussian transmission integral that was implemented as a precalculated lookup table.

Considering multi-beam linear interference, we obtained a contrast visibility factor that takes into account the individual parameters of each incident beam. Thus, we treated the case that an arbitrary number of laser beams -- which arrive from different directions, are laterally displaced, and have different individual amplitudes and beam radii -- are superimposed on a detector plane. While not resolving the detailed spatial interference pattern, we obtained integrated results as measured by a spectrometer or power meter.

Laser pulses are modified by passing through GOEs. We used the response-function formalism to describe linear non-resonant dispersion, resonant absorption (using a Franck--Condon model), and second-order non-resonant response. The latter allowed inclusion of sum-frequency and second-harmonic generation. While such processes had been treated extensively in the literature, we derived simplified approximate results that nevertheless take into account the Gaussian beam overlap between generating pulses and the Gaussian beam curvatures (i.e., focusing properties), and how both properties translate to the nonlinearly generated signal beam.

We derived spatial properties of Gaussian laser beams for graphical representation. We found the minimum number of vertices of a polygon mesh that keeps the local relative error below a given threshold. Such an object can be scaled to represent Gaussian beams with any other spatial parameters, and we derived the scaling factors. For color visualization, we also obtained the relation between the physical laser spectrum and the computer-graphics RGB(A) color system.

The model described in the current work forms the core of an interactive and immersive virtual-reality (VR) simulation of an ultrafast laser laboratory that we have recently developed \cite{brixner_femtopro_2023, noauthor_femtopro_nodate}. The results of the present work are also applicable, though, for other optics or spectroscopy simulations outside of femtoPro and beyond a VR context.

\begin{backmatter}
\bmsection{Funding}
This work was supported by the Fonds der Chemischen Industrie (FCI). We further acknowledge funding by the Julius-Maximilians-Universität Würzburg Project WueDive via Stiftung Innovation in der Hochschullehre.

\bmsection{Author contributions}
T.B.~derived the equations, coded the physical model, and wrote the paper. S.M.~coded didactic missions, carried out and analyzed the simulations, and prepared the figures. A.M.~coded the femtoPro core software and framework. T.B.~and S.v.M.\ supervised the project. All authors discussed the results and edited the manuscript.

\bmsection{Acknowledgment}
We thank Pavel Mal\'{y} for deriving equations of Gaussian field overlap and nonlinear signal-beam curvature. 

\bmsection{Disclosures}
``femtoPro'' is a registered trademark of the University of W\"urzburg. The software is available via the femtoPro website \cite{noauthor_femtopro_nodate}.

\bmsection{Data Availability Statement}
The data that support the findings of this study are openly available in Zenodo at \url{https://doi.org/10.5281/zenodo.15535423} \cite{brixner_dataset_2025}.

\bmsection{Supplemental document}
See the Supplementary Information for supporting content.
\end{backmatter}

%%%%%%%%%%%%%%%%%%%%%%% References %%%%%%%%%%%%%%%%%%%%%%%%%

%%%%%%%%%% If using BibTeX:
\bibliography{femtopro}

\begin{thebibliography}{10}
\newcommand{\enquote}[1]{``#1''}

\bibitem{spencer_general_1962}
G.~H. Spencer and M.~V. R.~K. Murty, \enquote{General ray-tracing procedure,}
  {\protect\JournalTitle{J. Opt. Soc. Am.}} \textbf{52}, 672--678 (1962).

\bibitem{noauthor_list_nodate}
\enquote{List of ray tracing software,}
  https://en.wikipedia.org/wiki/List\_of\_ray\_tracing\_software.

\bibitem{garcia_de_abajo_interaction_1999}
F.~J. {Garc{\'i}a de Abajo}, \enquote{Interaction of radiation and fast
  electrons with clusters of dielectrics: {{A}} multiple scattering approach,}
  {\protect\JournalTitle{Phys. Rev. Lett.}} \textbf{82}, 2776 (1999).

\bibitem{garcia_de_abajo_retarded_2002-1}
F.~J. {Garc{\'i}a de Abajo} and A.~Howie, \enquote{Retarded field calculation
  of electron energy loss in inhomogeneous dielectrics,}
  {\protect\JournalTitle{Phys. Rev. B}} \textbf{65}, 115418 (2002).

\bibitem{oskooi_meep_2010}
A.~F. Oskooi, D.~Roundy, M.~Ibanescu, \emph{et~al.}, \enquote{Meep: {{A}}
  flexible free-software package for electromagnetic simulations by the
  {{FDTD}} method,} {\protect\JournalTitle{Comput. Phys. Commun.}}
  \textbf{181}, 687--702 (2010).

\bibitem{comsol_comsol_nodate}
COMSOL, \enquote{{{COMSOL}}: {{Multiphysik-Software}} zur {{Optimierung}} von
  {{Designs}},} https://www.comsol.de.

\bibitem{lumerical_high-performance_nodate}
Lumerical, \enquote{High-performance photonic simulation software,}
  https://www.lumerical.com.

\bibitem{kristensen_modeling_2020}
P.~T. Kristensen, K.~Herrmann, F.~Intravaia, and K.~Busch, \enquote{Modeling
  electromagnetic resonators using quasinormal modes,}
  {\protect\JournalTitle{Adv. Opt. Photon.}} \textbf{12}, 612--708 (2020).

\bibitem{abramavicius_coherent_2009}
D.~Abramavicius, B.~Palmieri, D.~V. Voronine, \emph{et~al.}, \enquote{Coherent
  multidimensional optical spectroscopy of excitons in molecular aggregates;
  {{Quasiparticle}} versus supermolecule perspectives,}
  {\protect\JournalTitle{Chem. Rev.}} \textbf{109}, 2350--2408 (2009).

\bibitem{mitric_laser-field-induced_2009}
R.~Mitri{\'c}, J.~Petersen, and V.~{Bona{\v c}i{\'c}-Kouteck{\'y}},
  \enquote{Laser-field-induced surface-hopping method for the simulation and
  control of ultrafast photodynamics,} {\protect\JournalTitle{Phys. Rev. A}}
  \textbf{79}, 053416 (2009).

\bibitem{mancal_tmancal74quantarhei_2025}
T.~Man{\v c}al, \enquote{Tmancal74/quantarhei,}
  https://github.com/tmancal74/quantarhei (2025).

\bibitem{rose_efficient_2021}
P.~A. Rose and J.~J. Krich, \enquote{Efficient numerical method for predicting
  nonlinear optical spectroscopies of open systems,} {\protect\JournalTitle{J.
  Chem. Phys.}} \textbf{154}, 034108 (2021).

\bibitem{kenneweg_qdt_2024}
T.~Kenneweg, S.~Mueller, T.~Brixner, and W.~Pfeiffer, \enquote{{{QDT}} --- {{A
  Matlab}} toolbox for the simulation of coupled quantum systems and coherent
  multidimensional spectroscopy,} {\protect\JournalTitle{Comput. Phys.
  Commun.}} \textbf{296}, 109031 (2024).

\bibitem{brixner_femtopro_2023}
T.~Brixner, S.~Mueller, A.~M{\"u}ller, \emph{et~al.}, \enquote{{{femtoPro}}:
  Virtual-reality interactive training simulator of an ultrafast laser
  laboratory,} {\protect\JournalTitle{Appl. Phys. B}} \textbf{129}, 78 (2023).

\bibitem{noauthor_femtopro_nodate}
\enquote{{{femtoPro}},} https://www.femtopro.com.

\bibitem{han_assessing_2022}
C.~Han, G.~Xu, X.~Zheng, \emph{et~al.}, \enquote{Assessing the {{Effect}} of
  the {{Refresh Rate}} of a {{Device}} on {{Various Motion Stimulation
  Frequencies Based}} on {{Steady-State Motion Visual Evoked Potentials}},}
  {\protect\JournalTitle{Front. Neurosci.}} \textbf{15}, 757679 (2022).

\bibitem{wang_effect_2023}
J.~Wang, R.~Shi, W.~Zheng, \emph{et~al.}, \enquote{Effect of {{Frame Rate}} on
  {{User Experience}}, {{Performance}}, and {{Simulator Sickness}} in {{Virtual
  Reality}},} {\protect\JournalTitle{IEEE Trans. Vis. Comput. Graph.}}
  \textbf{29}, 2478--2488 (2023).

\bibitem{klein_optics_1986}
M.~V. Klein and T.~E. Furtak, \emph{Optics} (Wiley, Hoboken, 1986), 2nd ed.

\bibitem{born_principles_1999}
M.~Born and E.~Wolf, \emph{Principles of {{Optics}}: {{Electromagnetic Theory}}
  of {{Propagation}}, {{Interference}} and {{Diffraction}} of {{Light}}}
  (Cambridge University Press, Cambridge, 1999), 7th ed.

\bibitem{zinth_optik._2005}
W.~Zinth and U.~Zinth, \emph{{Optik. Lichtstrahlen -- Wellen -- Photonen}}
  (Oldenbourg Wissenschaftsverlag, M{\"u}nchen, 2005).

\bibitem{novotny_principles_2012}
L.~Novotny and B.~Hecht, \emph{Principles of Nano-Optics} (Cambridge University
  Press, Cambridge, 2012), 2nd ed.

\bibitem{boyd_nonlinear_2008}
R.~W. Boyd, \emph{Nonlinear Optics} (Academic Press, Burlington, 2008), 3rd ed.

\bibitem{saleh_fundamentals_2019}
B.~E.~A. Saleh and M.~C. Teich, \emph{Fundamentals of {{Photonics}}} (Wiley,
  Hoboken, 2019), 3rd ed.

\bibitem{mukamel_principles_1995}
S.~Mukamel, \emph{Principles of Nonlinear Optical Spectroscopy} (Oxford
  University Press, New York, 1995), 1st ed.

\bibitem{diels_ultrashort_2006}
J.-C. Diels and W.~Rudolph, \emph{Ultrashort Laser Pulse Phenomena:
  Fundamentals, Techniques, and Applications on a Femtosecond Time Scale}
  (Academic Press Inc, Amsterdam, 2006), 2nd ed.

\bibitem{trebino_frequency-resolved_2002}
R.~Trebino, \emph{Frequency-Resolved Optical Gating: {{The}} Measurement of
  Ultrashort Laser Pulses} (Springer, New York, 2002), 1st ed.

\bibitem{wollenhaupt_femtosecond_2007}
M.~Wollenhaupt, A.~Assion, and T.~Baumert, \enquote{Femtosecond laser pulses:
  {{Linear}} properties, manipulation, generation and measurement,} in
  \emph{Springer {{Handbook}} of {{Lasers}} and {{Optics}},}  F.~Tr{\"a}ger,
  ed. (Springer Science+Business Media, New York, 2007), pp. 937--983.

\bibitem{weiner_ultrafast_2009}
A.~M. Weiner, \emph{Ultrafast {{Optics}}} (John Wiley \& Sons Inc., Hoboken,
  2009), 1st ed.

\bibitem{cho_two-dimensional_2009}
M.~Cho, \emph{Two-{{Dimensional Optical Spectroscopy}}} (CRC Press, Boca Raton,
  2009).

\bibitem{hamm_concepts_2011-1}
P.~Hamm and M.~Zanni, \emph{Concepts and Methods of {{2D}} Infrared
  Spectroscopy} (Cambridge University Press, New York, 2011), 1st ed.

\bibitem{valkunas_molecular_2013}
L.~Valkunas, D.~Abramavicius, and T.~Man{\v c}al, \emph{Molecular {{Excitation
  Dynamics}} and {{Relaxation}}} (Wiley-VCH, Weinheim, 2013), 1st ed.

\bibitem{yuen-zhou_ultrafast_2014}
J.~{Yuen-Zhou}, J.~J. Krich, I.~Kassal, \emph{et~al.}, \emph{Ultrafast
  {{Spectroscopy}}} (IOP Publishing, Bristol, 2014), 1st ed.

\bibitem{muller_graph-based_2025}
A.~M{\"u}ller, S.~Mueller, T.~Brixner, and S.~{von Mammen}, \enquote{A
  {{Graph-Based Laser Path Solver Algorithm}} for {{Virtual Reality Laboratory
  Simulations}},} {\protect\JournalTitle{Simul. Model. Pract. Theory}} p.
  submitted (2025).

\bibitem{worku_propagation_2019}
N.~G. Worku and H.~Gross, \enquote{Propagation of truncated {{Gaussian}} beams
  and their application in modeling sharp-edge diffraction,}
  {\protect\JournalTitle{J. Opt. Soc. Am. A}} \textbf{36}, 859--868 (2019).

\bibitem{gatland_thin_2002}
I.~R. Gatland, \enquote{Thin lens ray tracing,} {\protect\JournalTitle{Am. J.
  Phys.}} \textbf{70}, 1184--1186 (2002).

\bibitem{stockman_cone_2019}
A.~Stockman, \enquote{Cone fundamentals and {{CIE}} standards,}
  {\protect\JournalTitle{Curr. Opin. Behav. Sci.}} \textbf{30}, 87--93 (2019).

\bibitem{delong_frequency-resolved_1994}
K.~W. DeLong, R.~Trebino, J.~Hunter, and W.~E. White,
  \enquote{Frequency-resolved optical gating with the use of second-harmonic
  generation,} {\protect\JournalTitle{J. Opt. Soc. Am. B}} \textbf{11},
  2206--2215 (1994).

\bibitem{delong_frog_2004}
K.~W. DeLong, \enquote{{{FROG}} 3.2.2,}  (2004).

\bibitem{brixner_dataset_2025}
T.~Brixner, S.~Mueller, A.~M{\"u}ller, and S.~{von Mammen}, \enquote{Dataset
  for {{femtoPro}}: {{Real-time}} linear and nonlinear optics simulations,}
  https://doi.org/10.5281/zenodo.15535423 (2025).

\end{thebibliography}

%%%%%%%%%% If preparing manually:
% \begin{thebibliography}{1}
% \newcommand{\enquote}[1]{``#1''}

% \bibitem{Zhang:14}
% Y.~Zhang, S.~Qiao, L.~Sun, Q.~W. Shi, W.~Huang, L.~Li, and Z.~Yang,
%   \enquote{Photoinduced active terahertz metamaterials with nanostructured
%   vanadium dioxide film deposited by sol-gel method,}
%   {\protect\JournalTitle{Optics Express}} \textbf{22}, 11070--11078 (2014).

% \bibitem{Optica}
% {Optica}, \enquote{{Optica Publishing Group},}
%   \url{http://www.opg.optica.org}.

% \bibitem{FORSTER2007}
% P.~Forster, V.~Ramaswamy, P.~Artaxo, T.~Bernsten, R.~Betts, D.~Fahey,
%   J.~Haywood, J.~Lean, D.~Lowe, G.~Myhre, J.~Nganga, R.~Prinn, G.~Raga,
%   M.~Schulz, and R.~V. Dorland, \enquote{Changes in atmospheric consituents and
%   in radiative forcing,} in \enquote{Climate Change 2007: The Physical Science
%   Basis. Contribution of Working Group 1 to the Fourth Assesment Report of
%   Intergovernmental Panel on Climate Change,}  S.~Solomon, D.~Qin, M.~Manning,
%   Z.~Chen, M.~Marquis, K.~B. Averyt, M.~Tignor, and H.~L. Miler, eds.
%   (Cambridge University Press, 2007).

% \end{thebibliography}

\includepdf[pages=-]{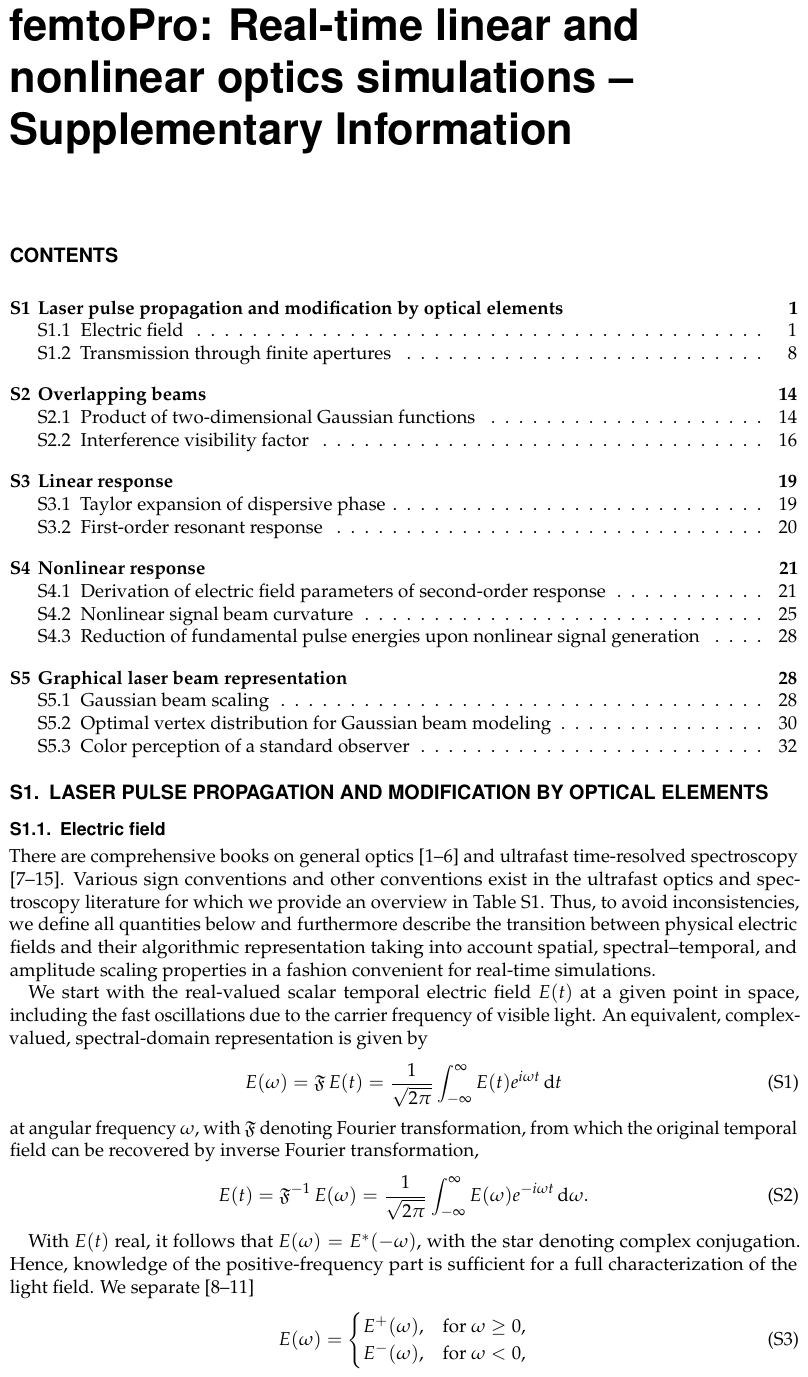}

\end{document}